\documentclass[namedreferences]{solarphysics}
\usepackage[optionalrh]{spr-sola-addons} % For Solar Physics
\usepackage{graphicx}        % For eps figures, newer & more powerfull
\usepackage{color}           % For color text: \color command
\usepackage{url}             % For breaking URLs easily trough lines
\newcommand{\etal}{{\it et al.}}

\newcommand{\adv}{    {\it Adv. Space Res.}}

\newcommand{\aap}{    {\it Astron. Astrophys.}}

\newcommand{\aapr}{   {\it Astron. Astrophys. Rev.}}

\newcommand{\apj}{    {\it Astrophys. J.}}
\newcommand{\apjl}{   {\it Astrophys. J. Lett.}}

\newcommand{\nat}{    {\it Nature}}

\newcommand{\solphys}{{\it Solar Phys.}}

\begin{document}

\begin{article}

\begin{opening}

\title{Explosive events - swirling transition region jets}

\author{W.~\surname{Curdt}$^{1}$\sep
        H.~\surname{Tian}$^{2}$\sep
        S.~\surname{Kamio}$^{1}$
       }
\runningauthor{Curdt et al.}
\runningtitle{Swirl in explosive events}

   \institute{$^{1}$ Max Planck Institute for Solar System Research, \\
   37191 Katlenburg-Lindau, Germany \\
                     email: \url{curdt@mps.mpg.de} \\
              $^{2}$ High Altitude Observatory (NCAR), \\
              Boulder, CO 80307, USA \\
                     email: \url{htian@hao.edu} \\
             }

\begin{abstract}

In this paper, we extend our earlier work to provide additional evidence for an alternative
scenario to explain the nature of so-called `explosive events'. The bi-directed,
fast Doppler motion of explosive events observed spectroscopically in the transition region
emission is classically interpreted as a pair of bidirectional jets moving upward
and downward from a reconnection site. We discuss the problems of such a model.
In our previous work, we focused basically on the discrepancy of fast Doppler motion
without detectable motion in the image plane. We now suggest an alternative
scenario for the explosive events, based on our observations of spectral line tilts
and bifurcated structure in some events. Both
features are indicative of rotational motion in narrow structures. We explain
the bifurcation as the result of rotation of hollow cylindrical structures
and demonstrate that such a sheath model can also be applied to explain the nature
of the puzzling `explosive events'. We find that the spectral tilt, the lack of
apparent motion, the bifurcation, and a rapidly growing number of direct observations
support an alternative scenario of linear, spicular-sized jets with a strong spinning motion.

\end{abstract}

\end{opening}
%-------------------------------------------------

\section{Introduction}
     \label{S-Introduction}

The term `explosive event' (EE) was first introduced by Dere, Bartoe, and Brueckner (1984).
Typically, EEs are characterized by their paired Doppler flows of $\pm$(50 - 150)~km~s$^{-1}$ in
opposite directions, with duration of one to three minutes, and small extent of
normally 1500~km to 2500~km. They often show up in bursts with a birth rate
of 2500 per second over the entire Sun, resulting in a huge number of 30,000 at
any one time on disk (Teriaca {\etal} 2004). They are best observed in emission
lines formed at transition region temperatures.
Recently, we reported a spectroscopic evidence of helicity in explosive events.
In the previous investigation (cf., Curdt and Tian 2011, hereafter referred to as CT1)
we basically used the extreme lack of an apparent motion in two EEs as argument against bi-directional
linear jets and suggested alternatively that a helical motion could be responsible
for the pairs of oppositely directed motions. We now find additional evidence in support
of such a scenario using arguments that were communicated over the years and
recent results obtained from spectroscopic observations by SoHO-SUMER and {\it Hinode}-EIS.
We suggest that the phenomenon `explosive event' is nothing else than the
spectroscopic signature of the rotational motion in a swirling jet.
Our argument is based on cases of observations of tilted spectral lines,
lack of apparent motion of the features in image sequences, bifurcated structure of
jets, and other observational features.

Direct observation of spinning motion is only possible if the rotation axis
does not coincide with the symmetry axis or if substructures of
an object can be resolved (e.g., Figure~3 in CT1). Pike and Mason (1998) have
presented SoHO-CDS observations of what they call tornados and suggest that
these are larger and spatially resolved versions of rotating H$\alpha$ spicules.
A multi-spacecraft observation of a rotating polar jet was communicated by
Kamio {\etal} (2010, hereafter referred to as KC1).
Recently, Wedemeyer-B\"ohm and van der Voort (2009) reported disk observations of `chromospheric
swirls' in CRISP Ca\,{\sc ii}~IR images. In all these cases the objects extended over tens of
arcseconds. However, spinning motion down to scales near the resolution of modern
space- or ground-based telescopes has recently also been reported by many observers
(e.g., Nistic\`o {\etal} 2009, Liu {\etal} 2011, Shen {\etal} 2011).
Suematsu {\etal} (2008) have noted the oscillatory behaviour with a
period of 1-1.5~min in {\it Hinode}-SOT images and have interpreted it as
the spinning motion of the spicule as a rigid body.

In analogy with distant stars that are spectroscopic binaries, spectroscopes are
able to detect rotation even for sub-resolution structures.
A spinning motion of a narrow feature will leave a typical signature in spectroscopic
data, even at sub-resolution scales. Normally, the dispersion direction is
orthogonal to the slit direction. However, in spectra of rotating fine structures
that cross the spectrometer slit at some angle the redshift and blueshift are
inclined relative to the dispersion direction, an effect often called 'spectral tilt'.
The idea that those small scale dynamical events that were later called EEs
might be nothing else than spinning spicular features was first brought up
by Pasachoff, Noyes, and Beckers (1968) which was based on H$\alpha$ spectra that
show spectral tilt. Beckers (1964) and Zirin (1966) have already reported
such an effect. Little later, Rompolt (1975) analysed systematically the
effects of rotational and expansional motions in filamentary structures on
spectrally resolved line profiles and interpreted the effects seen in H$\alpha$
spectra as indication of a well-ordered macroscopic motion in spicules and other
narrow solar structures. A decade later, such spectral tilt was also seen
in transition region emission (Cook {\etal} 1984) from the observation of symmetrical
Doppler flows in C\,{\sc iv} spectra obtained during a HRTS rocket flight.
Also in the SUMER spectra of EEs it was often found that the redshift is offset
from the blueshift along the slit by several pixels (Innes {\etal} 1997).

High-resolution Ca\,{\sc ii} images of {\it Hinode}-SOT have revealed that many fine structures have
a double-stranded nature with a separation of a few tenth of an arcsec that seems
to expand with time and limb ascent (Suematsu {\etal} 2008). Sterling, Harra, and Moore
(2010) have interpreted the double structure in spicules as a splitting process.

Despite their high velocities and duration of several minutes, EEs are almost
stationary, i.e., the redshift and the blueshift both stay in the
area that is imaged by the slit. For details we refer to the discussion in CT1.
This lack of apparent motion has already been
mentioned by Dere, Bartoe, and Brueckner (1989) and is not yet explained.

\begin{figure} %figure 1: tilt
 % \plotone{carton.eps}
  \includegraphics[width=9cm]{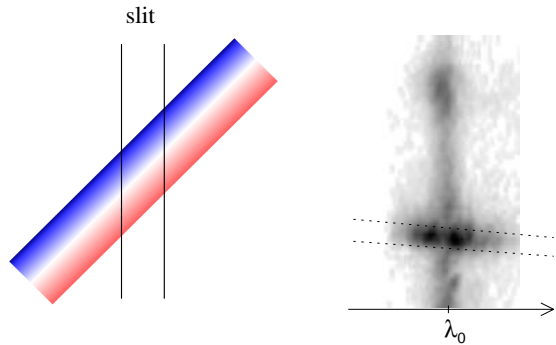}
  \caption {Spectrally resolved emission of Si\,{\sc iii} (120.6 nm) during an EE (right) is
  marked by the dotted lines. The event reaches flows from -40 to +40 km~s$^{-1}$
  if scaled as Doppler motion. The inclination angle between the Doppler flow and
  the dispersion direction can be explained by the rotation of a narrow,
  slit-size feature that -- if projected into the image plane -- is tilted by an
  angle $\alpha$ (left): the upflow (blue) is observed further up compared to the
  downflow (red). Even for sub-resolution features the spectral tilt provides an
  evidence of rotation.
  }
\end{figure}

\section{Observations and results} %%%%%%%%%%%%%%%%%%%%%%%%%%%%%%%%%%%%%%%%
      \label{S-general}

We focus on observations carried out on Nov 16 and 19, 2010 (cf., CT1) and on Nov 4, 2007 (cf., KC1)
that have already been described elsewhere. For observational details and the full description
of the results we refer to the original papers (cf., CT1 and KC1).
Here, we highlight particular details, namely the spectral tilt in the case of CT1 and
the bifurcation in the case of KC1. The recognition of these features
has not been the focal point of the original works and the discussion in a different
and new context goes beyond what was previously presented (cf., CT1 and KC1).

\subsection{Doppler flow without apparent motion}

In CT1 we have described a sit and stare study that was completed in the active regions
11124 and 11126. The spectra of the Si\,{\sc iii} emission line at 120.6 nm were
recorded on several days at a cadence of 10~s. In the data sets, small scale brightenings with
high Doppler motion were present and two of these typical EEs were discussed.
The empirical results of CT1 can be summarized as follows. In both cases the EE
started as if it were switched on from one exposure to the next. The line
brightened by a factor of $\ge 20$ and the emission
line split symmetrically with only little emission at rest, red- and blueshift
were seen at the same time, and there was no sign of acceleration in the
Doppler flow of up to $\approx$40~km~s$^{-1}$. The flow pattern did not change over the
event duration of $\approx$4~minutes, but there were two peaks in the lightcurve
that indicated a period of 150~s if interpreted as an oscillatory phenomenon.

In CT1 we have argued the extreme lack of apparent motion.
If both the upflow and the downflow stayed unchanged over four minutes within
the narrow spectrometer slit -- or in other words -- any lateral movement
along or across the slit exceeding 500~km had to be excluded, then a special
geometry would be required where both jets stayed within a cone angle of less
than a degree. In addition, this narrow cone had to be aligned with the line of sight.
This is highly unlikely. Although the problem of lacking apparent motion has already
been mentioned (cf., Dere, Bartoe, and Brueckner 1989, Innes 2004), yet it remained unexplained.
The discussion of this discrepancy has been the focal point of CT1 and the
argument is mentioned here only for the sake of completeness.

\subsection{Spectral tilt} %%%%%%%%%%%%%%
  \label{Spetral tilt}

A closer inspection of individual spectra has revealed another interesting
detail. In both data sets the Doppler flow of the EE is definitely inclined
relative to the dispersion direction. The spectral tilt is clearly visible
during the entire event. In the example shown in Figure~1 the linear offset
between the extremes of red- and blueshifted emission, {\small $\Delta$}$s$, is $\approx$ 2~px,
which corresponds to 1450~km. Let us assume a geometry with an angle between slit
and projected spin axis, $\alpha$, of 45$^\circ$. For such a case a cylinder diameter, $d$, of
$d$ = {\small $\Delta$}$s$ / sin $\alpha$ = 1000~km would be required,
appropriate for a decent spicule. Different from many other observations,
where only offset blobs of red- and blueshifted emission are seen, here all
velocities are present. We use this as a strong argument for rotation of a cylindrical shell,
since we only observe relatively little emission from the ions at rest.

From a review of published work on EEs we conclude that the spectral tilt reported here
is not an exceptional, but a rather common feature. In all three
cases presented by Innes {\etal} (1997) the blue wing is offset from the red wing.
Spectral tilt is also very obvious in the data analysed by Medoza-Torres {\etal} (2009, cf., Figure~2)
and Brkovi\'c and Peter (2004, cf., Figures~2 and 3), but has not been interpreted
as spinning motion previously.

\subsection{Bifurcation as a result of rotation } %%%%%%%%%%%%%%
  \label{bifurcation}

In KC1 we have presented a polar jet that was observed in a coordinated campaign by
{\it Hinode}-XRT, {\it Hinode}-SOT, STEREO-EUVI, and by the spectrometers
SoHO-SUMER and {\it Hinode}-EIS. EIS obtained a sequence of rasters in
Fe\,{\sc xii} (19.5~nm) and He\,{\sc ii} (25.6~nm), while SUMER scanned the same
polar region in Ne\,{\sc viii} (77.0~nm) and in O\,{\sc iv} (79.0~nm).
This multi-spacecraft observation is a fortuitous and rare case of two
spectrometers pointing their slit during a raster timely to the right location to observe
such a short-lived phenomenon. In Figure~2 we show the jet as seen by XRT (Al\_poly, left)
and an image in He\,{\sc ii} (30.4 nm, middle). The position of the EIS and SUMER slits -- both rastering
inward but in reverse direction -- is indicated in the XRT image in yellow and magenta, respectively.
The redshift of +50~km~s$^{-1}$ seen by EIS in He\,{\sc ii} and the blueshift of 120 km~s$^{-1}$
seen on the opposite side by SUMER in O\,{\sc iv} reveal the clockwise rotation in the
periphery of the jet. In emission of plasma at temperatures of $\approx$1~MK
as seen by XRT, the jet appears bifurcated. The bright streaks at the periphery
of the jet are close to the locations where the spectroscopes observe the Doppler flow.
This suggests that those bright {\it strands} are the edges of a rotating
cylindrical structure, and are bright in optically-thin soft X-ray
emission due to the longer path-lengths through the cylinder at its edges
compared to the {\it body} of the cylinder. The rotational motion of the structure
also explains the lack of lateral movement in the image plane.

The dotted box in the XRT image indicates the FOV observed by SOT in Ca\,{\sc ii}~H
(Tsuneta {\etal} 2008) as shown in the right panel.
There is indeed a chromospheric counterpart of the SXR and EUV features, the
tilted strands at the base of the jet seem to be a part of the helical structure.

We note that the jet described in KC1 does not propagate along a strict radial
trajectory. We further note that the two branches of the jet are not entirely
parallel and differ in strength. Wiggles or kinks along the trajectory (Beckers 1972),
and periodicities in the lightcurve (CT1) seem to mimic asymmetries and indicate
a spiraling or swirling motion rather than a sheer solid body rotation.

\begin{figure} %figure 2: XRT_EUVI
 % \plotone{carton.eps}
  \includegraphics[width=12cm]{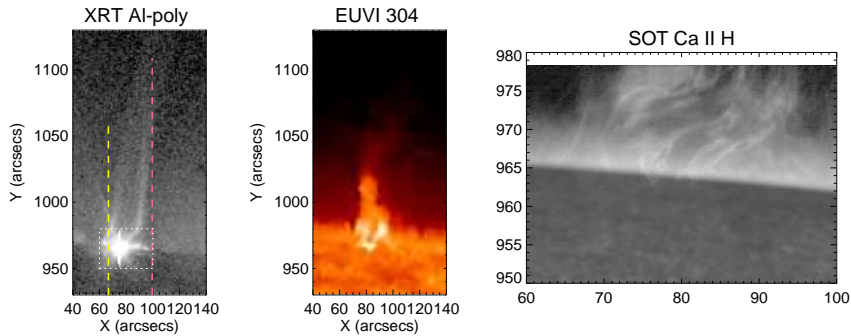}
  \caption {Polar jet as observed by {\it Hinode}-XRT (left), {\it Hinode}-SOT (right),
  and
  \mbox{STEREO-}EUVI (middle). The bifurcation seen in the XRT-jet is definitively
  associated with a rotational movement, since the spectrometers SUMER (magenta) and EIS (yellow)
  observe large Doppler flows in opposite directions at the marked slit positions (adapted from KC1).
  The Ca\,{\sc ii}~H image corresponds to the dotted box in the XRT-image and
  proves that the jet has indeed a chromospheric counterpart and
  is rooted in the chromosphere.}
\end{figure}

Following the bifurcation argument, we have carried out a simple numerical experiment and
have assumed a spicular feature that rotates with a tangential velocity of 60~km~s$^{-1}$.
The spicule is observed with an aspect angle
between spin axis and line-of-sight, $\delta$, that varies from 0$^\circ$ to 90$^\circ$.
In addition, we assume that 5\% of the emission comes from upward motion.
In other words -- we assume a spiraling motion. Such a faint upward contribution
of a few percent of the total emission has recently been suggested, based on the
analysis of blueward asymmetries in TR and coronal lines (e.g., Tian {\etal} 2011).
In Figure~3 we show the calculated spectral
profiles that should appear in SUMER spectra in different geometries.

\begin{figure} %figure 3: cartoon
 % \plotone{carton.eps}
  \includegraphics[width=\textwidth]{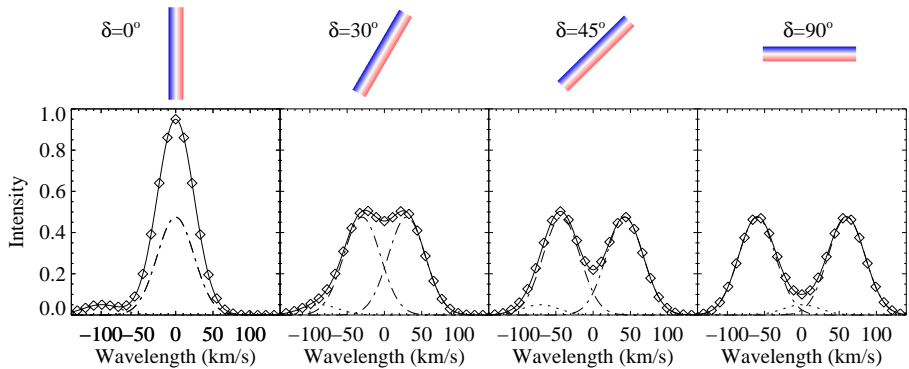}
  \caption {Simplified model of a bi-stranded, rotating jet for different aspect
  angles $\delta$ that can reproduce the SUMER spectra of EEs reported in
  CT1. In CT1 the model assumes a contribution of 47.5\% for the red and blue strands
  (dot-dashed) and of 5\% from the longitudinal component (dotted). The solid curve
  shows the summed emission. The 45$^\circ$ case of the summed emission is in
  good agreement with the observations (adapted from CT1).
  }
\end{figure}

\section{Discussion}

In {\it Hinode}-SOT Ca\,{\sc ii} filtergraph data spicules often show a double structure
with a separation of less than an arcsec. In an initial study, Suematsu {\etal} (2008)
have reported that over 50 \% of spicules they observed show a double structure.
They have concluded that
such double-threaded features clearly seen in their Figures~2 and 3 must be spinning as a rigid body.
Similarly, Sterling {\etal} (2010) suspect that the double structure in spicular features is
the result of a spinning motion. They also suggest a concept of
spicules being miniature EUV jets that feed the corona. This is in line with
the observation of a double-threaded X-ray jet associated with the rotating
macrospicule shown in Figure~2. This event is much larger than the events discussed in CT1.
We show it, since in this case the bifurcation can definitively be associated with the
rotation of the jet's hot skin. We suggest -- similar to Pike and Mason (1998),
Innes (2004) and others -- that this scenario can be scaled down to sizes at or below the resolution
limit of today's instruments and should be interpreted as a larger version of the
same phenomenon.  We are not aware of any physical reason for a lower limit of
spicular features, which should allow us to interpret our observation as if we
were observing with a magnifying glass and to suggest that they also rotate at
subresolution scales, although we have no conclusive evidence for this.

Recently, Beck and Rezaei (2011) have identified in POLIS Ca\,{\sc ii}~H
off-limb spectra elongated structures that are often flanked all along their
extension by velocities in opposite direction. Pariat, Antiochos, and DeVore (2009) have
suggested a MHD model to reproduce the ejection of a fast moving helical jet.

Pasachoff, Noyes, and Beckers (1968) interpreted the spectral tilt in H$\alpha$ spectra
as differential mass motion on the sides of the features, i.e., as a peripheral
rotation, a notion that is also found in the review of Beckers (1972).
Budnik {\etal} (1998) have shown that spicules at their outer boundary
have a hotter outer envelope, which in consequence forms a cylindrical shell at
TR temperatures around a confined cool core.

In CT1, we have presented extreme cases of Doppler motion that does not show
a detectable signature of bulk motion along or across the slit.
In an attempt to solve the problem one can still argue that the jets may not be
visible at their full extent, and that visibility is mostly for the red and blue
footpoints near the reconnection site that continuously replenishes the pair of jets.
But even for such a scenario it is difficult to understand why EEs
show the red and the blue component at the same time in most cases.

The observational fact that the statistical dominance of the blue component
over the red component of EEs observed at disk center is reduced near the limb
(Innes and T\'oth 1999) would be a natural implication of the scenario presented here.

\section{Conclusion} %%%%%%%%%%%%%%%%%%%%%%%%%%%%%%%%%%%%%%%%
      \label{S-Conclusion}

The work presented as case study in CT1 has significantly been expanded, and
the bifurcation and spectral tilt have been added as further evidence that at least
some EEs are not due to bi-directional reconnection flows.
The detailed empirical model presented here, of the explosive events being
rotating cylindrically-shaped jets, is fully consistent with the
observations. Our new interpretation can also account for unexplained features
found in older literature.

The concept that the EEs are indeed a manifestation
of the swirling motion of a blowout jet has been substantiated by direct
empirical evidence from events at larger scales and by the observed spectral
tilt explaining the bifurcation as the result of a hotter sheath around a
cylindrical body. The existence of such a cylindrical shell at TR temperature
around a confined cool core is a direct consequence
of the observation of Budnik {\etal} (1998). The basic idea that this cylinder
is rotating is consolidated by well-established arguments. It can easily explain
the old problem of lacking apparent motion in paired red-
and blueshifted Doppler flows, the observed spectral tilt, and the bifurcation
or double-stranded nature of jets if observed at higher temperatures.
The fact that the flows in EEs are easily detectable by spectroscopes,
but often without a counterpart in images at lower temperature could be a direct
consequence of their sub-resolution nature at chromospheric temperatures and the
widening of the rotating sheath around the cool H$\alpha$ emitting core.
We follow the conclusions derived from H$\alpha$ spectroscopic observations
(Rompolt 1975, Pasachoff, Noyes, and Beckers 1968) that spicules must be small-scale and
often unresolved features rotating in the same way as macrospicules do.

The spectral tilt in EEs reported here and the lack of apparent motion
discussed in CT1 led us to the concept of spinning narrow features that, if
observed at chromospheric wavelengths, appear as spicules and show up in
TR spectra as EEs. Such a correspondence has long been postulated (Cook {\etal} 1984)
and explicitly been mentioned by Pasachoff, Jacobson, and Sterling (2009).
Sterling, Harra, and Moore (2010) postulate even an EUV and SXR counterpart of
chromospheric spicular features. Our example shown in Figure~2 clearly
demonstrates that this is indeed the case on a macrospicular scale.

The case study presented here does not discuss the processes that generate EEs
or spicules. It does not even provide conclusive evidence for a one-to-one
association between spicules and EEs. It is possible that only one class of spicules
may be spinning. And it may also be -- as has already been noted by
Innes (2004) -- that two classes of EEs exist. We plan a systematic analysis
of the spectral tilt and the red-blue asymmetry in different geometries to answer this
question, while noting that we may have to wait for IRIS, should it turn out
that SUMER data is not good enough to disentangle the spatio-temporal ambiguity.

%%%%%%%%%%%%%%%%%%%%%%%%%%%%%%%%%%%%%%%%%%%%%%%%%%%%%%%%%%%%%%%%%%%%%%%%%%%
%\appendix

%%%%%%%%%%%%%%%%%%%%%%%%%%%%%%%%%%%%%%%%%%%%%%%%%%%%%%%%%%%%%%%%%%%%%%%%%%%
\begin{acks}
The SUMER project is financially supported by DLR, CNES, NASA, and the ESA
PRODEX Programme. SUMER is part of SoHO of ESA and NASA.
H.T. is supported by the ASP Postdoctoral Fellowship Program of NCAR.
The National Center for Atmospheric Research is sponsored by the
National Science Foundation. We thank the anonymous referee and B.N. Dwivedi for
constructive comments that helped to improve the clarity of this communication.
\end{acks}

%\bibliographystyle{spr-mp-sola}

%\bibliography{helicity}

\end{article}

\end{document}